\documentstyle[12pt]{article}
\def\beq{\begin{equation}}
\def\eeq{\end{equation}}
\def\bea{\begin{eqnarray}}
\def\beaa{\begin{eqnarray*}}
\def\eea{\end{eqnarray}}
\def\eeaa{\end{eqnarray*}}
\def\bq{\begin{quote}}
\def\eq{\end{quote}}
\def\gappeq{\mathrel{\rlap {\raise.5ex\hbox{$>$}}
{\lower.5ex\hbox{$\sim$}}}}

\def\lappeq{\mathrel{\rlap{\raise.5ex\hbox{$<$}}
{\lower.5ex\hbox{$\sim$}}}}

\def\big1{\mbox{\boldmath$1$}}
\def\NP{{\it Nucl.Phys.} }
\def\PL{{\it Phys.Lett.} }

\begin{document}
\pagestyle{empty}
\begin{flushright}
CERN-TH/2000-013\\
math-th/0001---
\end{flushright}
\vspace*{1cm}

\begin{center}
{\bf SOLUTIONS OF $D_\alpha$ = 0  }\\
{\bf FROM HOMOGENEOUS INVARIANT FUNCTIONS}\\
\vspace*{1cm}

F. Buccella \footnote{On leave of absence from Dipartimento di Scienze Fisiche,
Universit\`a di Napoli.}\\
\vspace*{0.5cm}
Theoretical Physics Division, CERN\\
CH - 1211 Geneva 23\\
\vspace*{3cm}

Abstract
\end{center}

We prove that the existence of a  homogeneous invariant of degree $n$
for a representation of a semi-simple Lie group guarantees the
existence of non-trivial solutions of $D_{\alpha} = 0$: these correspond
to the maximum value of the square of the invariant divided by the norm
of the representation to the $n^{th}$ power.
\vspace*{8cm}
\begin{flushleft}
CERN-TH/2000-013\\
January 2000
\end{flushleft}
\vfill\eject
\pagestyle{plain}
\setcounter{page}{1}
The search of solutions of the equation:
\beq
 D_{\alpha}=0
\label{one}
\eeq
is a necessary tool for the classification of non-trivial
supersymmetric vacua \cite{FZ}.
Several years ago a sufficient condition was proposed for
the validity of (\ref{one}), namely the existence of an invariant $F(z)$,
such that:
\beq
(\frac{\delta F}{\delta z_a})_{z_a} = k z_a^*
\label{two}
\eeq
with $k \not= 0$ \cite{BDFS}. On the basis of absence of known
counterexamples the condition has been conjectured to be also
necessary \cite{BDFS} and the proof has been given by Procesi and Schwarz
\cite{PS}.
Here we want to show that a sufficient condition for the existence of
non-trivial solutions of (\ref{one}) is the existence of a  homogeneous
invariant $F^n (z)$ of degree $n$.
 The proof goes as follows.\\ Be $F(z_a)$ a
homogeneous invariant of degree $n$; let us consider the function:
\beq
G(z_a,z_a^*)=\frac{F(z_a)F(z_a^*)}{{N(z_a})^n}
\label{three}
\eeq
where:
\beq
N(z_a)= \Sigma_a z_az_a^*~.
\label{four}
\eeq
Let us look for the maxima of $G$, which is real, positive and homogeneous
of degree 0 in the domain:
\beq
\frac{1}{2} \leq N(z_a)\leq 1~.
\label{five}
\eeq
A maximum certainly exists, since $G$ is a continuous function of the
real variables $x_a=\frac{z_a +z_a^*}{2}$ and $y_a=\frac{z_a-z_a^*}{2i}$
defined on a closed and limited set (Weierstrass theorem).
Since $G$ is a homogeneous function of degree 0, it is constant along the
intersection of the radii, which start from the origin, with the set defined
by (\ref{five}). Therefore the maximum will occur also on an internal point
$z_a^0$ of the domain, where:
\beq
\frac{\delta G}{\delta z_a}_{z_a=z_a^0}=
\left(\frac{
\frac{\delta F}{\delta z_a}F(z_a^*)
N(z_a)-nz_a^*F(z_a)F(z_a^*)}{N(z_a)^{n+1}}\right)_{z_a = z_a^0}
\label{six}
\eeq
should vanish. At that point, $F(z_a)$ and $F(z_a^*)$ are different from zero,
since the maximum of a positive function, which is not identically zero, is
positive. $N(z)$ is also positive, so from (\ref{six}) we get:
\beq
\frac{\delta F}{\delta z_a}_{z_a=z_a^0} =kz_a^{0*}
\label{seven}
\eeq
with $k=\frac{nF(z_a)}{N(z_a)}\not= 0$,
which guarantees \cite{BDFS} the vanishing of the $D$ term for $z_a^{(*)} =
z_a^{0(*)}$. To give some application
of the theorem just shown, let us consider an irreducible representation
$\phi$ of a semi-simple Lie group
$G$. We may decompose the symmetric product of two $\phi$'s along
irreducible representations of $G$:
\beq
(\phi \times \phi)_S =\Sigma_l\phi_l
\label{eight}
\eeq
The number of independent quartic invariants, bilinear in $\phi$ and its
complex conjugate $\phi^*$, is given by the number of terms in the r.h.s.
of (\ref{eight}). More precisely they are:
\beq
I_l = N(\phi \times \phi)_l~.
\label{nine}
\eeq
Also the invariant:
\beq
I_{\alpha} = N(\phi^* \times \phi)_{adjoint}~,    
\label{ten}
\eeq
which is proportional to $D_{\alpha} D^{\alpha}$, is a combination of the
$I_l$'s.
If the sum in (\ref{eight}) has only one term, there is only one independent
quartic invariant, bilinear in $\phi$ and $\phi^*$, $I_{\alpha}$ is
proportional to ${N(\phi)}^2$ and there is no non-trivial solution
to (\ref{one}). In that case we can conclude that we cannot write an invariant,
which depends only on $\phi$.
Let us now consider three cases, where there is a cubic invariant built
in terms of an irreducible complex representation of a simple group: according
to the theorem just shown, there will be solutions of (\ref{one}). We shall
consider the 6 of $SU(3)$, the 15 of $SU(6)$ and the 27 of $E(6)$ and their
symmetric products:
\bea
(6 \times 6)_S &=& 15 + \bar{6}  \nonumber   \\
(15 \times 15)_S &=& 105 + \bar{15} \nonumber \\
(27 \times 27)_S &=& 351 + \bar{27}  
\label{eleven}
\eea
So there are  two independent quartic invariants bilinear in the 6 (or the
15 or the 27) and in the $\bar{6}$ (or the $\bar{15}$ or the $\bar{27}$),
and one has:
\bea
18~ N(6 \times 6)_{\bar{6}} &+& 15~ N	(6\times\bar{6})_8 = 8~	{N(6)}^2
\nonumber \\
 9~ N(15 \times 15)_{\bar{15}} &+& 12	~ N(15 \times \bar{15})_{35} = 8~
{N(15)}^2
\nonumber
\\
15~ N(27 \times 27)_{\bar{27}} &+& 9~ N(27 \times \bar{27})_{78} = 2~
{N(27)}^2 
\label{twelve}
\eea
where the second terms in the l.h.s.'s of (\ref{twelve}) are just proportional
to $D_{\alpha}D^{\alpha}$. The two terms in the l.h.s.'s of (\ref{twelve}) have
the intriguing property that one vanishes when the other takes its
maximum. The vanishing of the second term when the second
takes its maximum is a consequence of the theorem we have just shown.
In fact, when $\phi_a$ is on a critical orbit of an
irreducible representation $\phi$, the invariants:
$$
{N~(6\times 6)_{\bar 6} \over N~(6)^2} \quad {\rm and}\quad 
{N~(6\times 6 \times 6)_1 \over N~(6)^3}
$$
are both proportional to $(C^{6~~~6~~\bar
6}_{\varphi_a~\varphi_a~\varphi_a^*})^2$, which implies that, 
if $\phi_a$ is a maximum
for the first one, it is a maximum  for the second one as well. 
  (The implication in the opposite direction does not hold for a
non-critical orbit, since in that case $N(\phi\times\phi)_{\phi^*}$ receives 
contributions also from $\phi^* \not=
\phi^*_a$.)  A similar property holds for the 15
of $SU(6)$ and the 27 of $E(6)$. The $D$ term
vanishes in the $SO(3)$, $Sp(6)$ or $F(4)$ invariant direction respectively.
From the other side the first terms in (\ref{twelve}) vanish in the $SU(2)$,
$SU(4)\times SU(2)$ or $SO(10)$ invariant direction, respectively. This is
not surprising, because, when $\varphi_a$ correspond to the maximal weight of
the representation, as in the cases just mentioned,
$(\varphi_a
\times \varphi_a)$ has components only along the representation with the
highest maximal weight. The necessity of the existence of an invariant for
the existence of solutions  of (\ref{one}), already proved in
\cite{PS}, implies that no invariant can be constructed with a representation
unable to supply non-trivial solutions to (\ref{one}). It applies to the case
with only one term in the r.h.s. of (\ref{eight}) and $D_{\alpha}D^{\alpha}$
proportional to ${N(\phi)}^2$, but also to cases with more than one term
present in (\ref{eight}). As an example, let us consider the 16 spinorial
representation of $SO(10)$, for which:
\bea   
&&(16 \times 16)_S = 126 + 10 \nonumber \\
&&32~ N(16 \times 16)_{10} + 16 ~N(16 \times \bar{16})_{45} = 5~ {N(16)}^2 ~.
\label{thirteen}
\eea
While the maximum of the second term in the l.h.s. of (\ref{thirteen}) in
the $SU(5)$ invariant direction corresponds to the vanishing of the
first term \cite{BRS}, the maximum of the first term in the $SO(7)$ invariant
direction corresponds to a minimum, but not to a vanishing value for
$D_\alpha D^\alpha$. In fact no $SO(10)$ invariant can be built only with a 16.

For semi-simple groups we shall consider the bifundamental 
$(N,\bar{M})$ representation of $SU(N) \times SU(M)$. For $M=N$, the existence
of the invariant:
\beq
\epsilon_{\beta_1 \ldots \beta_N}^{\gamma_1 \ldots \gamma_N}
\phi_{\alpha_1}^{\beta_1}\ldots\phi_{\gamma_N}^{\beta_N} 
\label{fourteen}
\eeq
implies the existence
 of a solution of (\ref{one}), namely the singlet under
the sum of the two $SU(N)$. 
If $N > M$ the contribution $\varphi_M^2$ to $D_\alpha D^\alpha$ from $SU(N)$
cannot vanish, since 
\beq
{(D_\alpha D^\alpha)_{SU(N)}\over g_N^2} > {(D_\alpha D^\alpha)_{SU(M)}\over g^2_M}
\label{fifteen}
\eeq
and the r.h.s. of (\ref{fifteen}) is non-negative: in fact no invariant can be
built in that case. For
$N = 3, M = 2$ one has
\beq
(\phi \times \phi)_s = (6,3) + (\bar 3,1)
\label{sixteen}
\eeq
and $N(\phi\times\phi)_{(\bar 3,1)}$ takes its maximum in the direction
where the contribution to $D_\alpha D^\alpha$ from $SU(2)$ (but not from
$SU(3)$) vanishes.

\vspace*{1cm}
\noindent
{\bf Acknowledgements}

Instructive and inspiring discussions with Prof. S. Ferrara are
gratefully acknowledged.

\end{document}